\documentclass{elsarticle}
\usepackage{lineno,hyperref,natbib}

\usepackage{amssymb}

\begin{document}

\begin{frontmatter}



\title{Band crossings in $^{168}$Ta: a particle-number conserving analysis}



\author{Zhen-Hua Zhang\fnref{contact}}
\ead{zhzhang@ncepu.edu.cn}
\author{Miao Huang}
\address{Mathematics and Physics Department,
              North China Electric Power University, Beijing 102206, China}

\begin{abstract}
The structures of two observed high-spin rotational bands in the
doubly-odd nucleus ${}^{168}$Ta are investigated using the cranked shell model
with pairing correlations treated by a particle-number conserving method,
in which the blocking effects are taken into account exactly.
The experimental moments of inertia and alignments
are reproduced very well by the calculations,
which confirms the configuration assignments for these two bands in previous works.
The backbending and upbending mechanisms in these two bands are analyzed in detail by calculating the
occupation probabilities of each orbital close to the Fermi surface
and the contributions of each orbital to the total angular momentum alignments.
The investigation shows that the level crossings in the first backbending
is the neutron $i_{13/2}$ crossing and the in second upbending
is the proton $h_{11/2}$ crossing.
\end{abstract}

\begin{keyword}
particle-number conserving method \sep
pairing correlation \sep
moment of inertia \sep
band crossing

\PACS 21.10.Re \sep 21.60.-n \sep 21.60.Cs \sep 27.70.+q

\end{keyword}

\end{frontmatter}


\section{Introduction}{\label{sec:intro}}

The high-spin rotational structures of the rare-earth nuclei with proton
$Z\approx72$ and neutron $N\approx94$ have drawn lots of attention due to
the existence of exotic excitation modes~\cite{Odegard2001_PRL86-5866,
Djongolov2003_PLB560-24,
Pattabiraman2007_PLB647-243,
Paul2007_PRL98-012501}.
Great efforts have been carried out to find these novel phenomena, {\it e.g.},
wobbling mode~\cite{Odegard2001_PRL86-5866,Bohr1975_Book,Jensen2002_PRL89-142503,
Amro2003_PLB553-197, Bringel2005_EPJA24-167,
Hartley2009_PRC80-041304R, Hartley2011_PRC83-064307}.
Meanwhile, a considerable amount of data of the high-spin rotational bands
in this mass region have been obtained in this process.
Especially some doubly-odd nuclei, {\it e.g.}, $^{166, 168, 170, 172}$Ta
\cite{Hartley2010_PRC82-057302, Wang2010_PRC82-034315,
Aguilar2010_PRC81-064317, Hojman2000_PRC61-064322},
$^{168, 170, 172, 174, 176}$Re~\cite{Hartley2016_PRC94-054329, Hartley2013_PRC87-024315,
Hartley2014_PRC90-017301, Guo2012_PRC86-014323, Cardona1999_PRC59-1298}, {\it etc.},
which are characterized by fairly small quadrupole deformations,
provide a good opportunity to the understanding
of the dependence of band crossing frequencies and angular momentum alignments on
the occupation of specific single-particle orbitals.

Furthermore, these data provide a benchmark for various nuclear models,
{\it e.g.}, the cranked Nilsson-Strutinsky
method~\cite{Andersson1976_NPA268-205},
the Hartree-Fock-Bogoliubov cranking model
with Nilsson~\cite{Bengtsson1979_NPA327-139}
and Woods-Saxon potentials~\cite{Nazarewicz1985_NPA435-397, Cwiok1987_CPC46-379},
the cranking non-relativistic~\cite{Dobaczewski1997_CPC102-166}
and relativistic mean-field models~\cite{Afanasjev1996_NPA608-107},
the projected shell model~\cite{Hara1995_IJMPE4-637},
the projected total energy surface approach~\cite{Tu2014_SCPMA57-2054}, {\it etc}.
For example, two high-spin rotational bands in the doubly-odd nuclei ${}^{168}$Ta
have been extended up to spin $\sim 40\hbar$ in Ref.~\cite{Wang2010_PRC82-034315},
in which the second band crossing in ${}^{168}$Ta has been observed for the first time.
However, the cranked shell model (CSM) can only reproduce
the first neutron crossing~\cite{Wang2010_PRC82-034315}.
The difficulty of the CSM in reproducing the second proton crossing
at high-spin region may come from a change of
deformation with increasing rotational frequency and/or the improper
treatment of pairing correlations~\cite{Hartley2005_PRC72-064325, Jensen2001_NPA695-3}.
Therefore, it is interesting to investigate the proton alignments
of $h_{11/2}$ orbitals involving
in this nucleus using a reliable nuclear model.

In the present work, the CSM with
pairing correlations treated by a particle-number conserving (PNC)
method~\cite{Zeng1983_NPA405-1, Zeng1994_PRC50-1388} will be used
to investigate the band crossings of the two high-spin rotational bands observed
in ${}^{168}$Ta~\cite{Wang2010_PRC82-034315}.
In contrary to the conventional Bardeen-Cooper-Schrieffer or
Hartree-Fock-Bogoliubov approaches, the Hamiltonian is solved directly
in a truncated Fock-space in the PNC method~\cite{Wu1989_PRC39-666}.
Therefore, the particle-number is conserved and the Pauli blocking effects are taken into account exactly.
The PNC-CSM has been employed successfully for describing various nuclear phenomena, {\it e.g.},
the odd-even differences in moments of inertia (MOIs)~\cite{Zeng1994_PRC50-746},
identical bands~\cite{Liu2002_PRC66-024320, He2005_EPJA23-217},
super-deformed bands~\cite{Liu2002_PRC66-024320, Xiang2018_CPC42-54105}
nuclear pairing phase transition~\cite{Wu2011_PRC83-034323},
antimagnetic rotation~\cite{Zhang2013_PRC87-054314, Zhang2016_PRC94-034305},
high-$K$ isomers in the rare-earth~\cite{Zhang2009_NPA816-19, Zhang2009_PRC80-034313,
Li2013_ChinPhysC37-014101, Zhang2016_SciChinaPMA59-672012}
and actinide nuclei~\cite{He2009_NPA817-45, Zhang2011_PRC83-011304R,
Zhang2012_PRC85-014324, Zhang2013_PRC87-054308, Li2016_SciChinaPMA59-672011}, {\it etc}.
The PNC scheme has also been adopted both in non-relativistic~\cite{Pillet2002_NPA697-141, Liang2015_PRC92-064325}
and relativistic mean-field models~\cite{Meng2006_FPC1-38}
and the total-Routhian-surface method with the
Woods-Saxon potential~\cite{Fu2013_PRC87-044319, Fu2013_SCPMA56-1423}.
Most recently, the shell-model-like approach, originally referred to as PNC method, based on the
cranking covariant density functional theory has been developed~\cite{Shi2018_PRC97-034317}.
Note that the covariant density functional theory provides
a consistent description of the nuclear properties,
especially the spin-orbital splitting,
the pseudo-spin symmetry~\cite{Arima1969_PLB30-517, Hecht1969_NPA137-129, Ginocchio1997_PRL78-436,
Meng1998_PRC58-628R, Meng1999_PRC59-154, Chen2003_CPL20-358,
Long2006_PLB639-242, Liang2011_PRC83-041301R, Lu2012_PRL109-072501, Liang2015_PR570-1}
and the spin symmetry in the anti-nucleon spectrum~\cite{Zhou2003_PRL91-262501, Liang2010_EPJA44-119},
and is reliable for the description of nuclei
far away from the $\beta$-stability line~\cite{Meng2006_PPNP57-470, Zhao2010_PRC82-054319}, etc.
Similar approaches with exactly conserved particle number when treating
the paring correlations can be found in Refs.~\cite{Richardson1964_NP52-221, Pan1998_PLB422-1,
Volya2001_PLB509-37, Jia2013_PRC88-044303, Jia2013_PRC88-064321, Chen2014_PRC89-014321}.

This paper is organized as follows.
A brief introduction to the PNC treatment of pairing correlations within
the CSM is presented in Sec.~\ref{sec:pnc}.
The calculated results for two observed high-spin rotational
bands in ${}^{168}$Ta using PNC-CSM are shown in Sec.~\ref{sec:resu}.
A brief summary is given in Sec.~\ref{sec:summ}.

\section{Theoretical framework}{\label{sec:pnc}}

The cranked shell model Hamiltonian of an axially symmetric
nucleus in the rotating frame can be written as
\begin{eqnarray}
 H_\mathrm{CSM}
 & = &
 H_0 + H_\mathrm{P}
 = H_{\rm Nil}-\omega J_x + H_\mathrm{P}
 \ ,
 \label{eq:H_CSM}
\end{eqnarray}
where $H_{\rm Nil}$ is the Nilsson Hamiltonian~\cite{Nilsson1969_NPA131-1},
$-\omega J_x$ is the Coriolis interaction with the cranking frequency $\omega$ about the
$x$ axis (perpendicular to the nuclear symmetry $z$ axis).
$H_{\rm P}$ is the pairing interaction,
\begin{eqnarray}
 H_{\rm P}(0)=
  -G_{0} \sum_{\xi\eta} a^\dag_{\xi} a^\dag_{\bar{\xi}}
                        a_{\bar{\eta}} a_{\eta}
  \ ,
\end{eqnarray}
where $\bar{\xi}$ ($\bar{\eta}$) labels the time-reversed state of a
Nilsson state $\xi$ ($\eta$),
and $G_0$ is the effective strength of monopole pairing interaction.

Instead of the usual single-particle level truncation in conventional
shell-model calculations, a cranked many-particle configuration (CMPC)
truncation is adopted, which is crucial
to make the PNC calculations for low-lying excited states both
workable and sufficiently accurate~\cite{Molique1997_PRC56-1795, Wu1989_PRC39-666}.
Usually a CMPC space with the dimension of 1000 will be enough for the investigation
of the rare-earth nuclei.
By diagonalizing the $H_\mathrm{CSM}$ in a sufficiently
large CMPC space, sufficiently accurate solutions for low-lying excited eigenstates of
$H_\mathrm{CSM}$ can be obtained, which can be written as
\begin{eqnarray}
 |\Psi\rangle = \sum_{i} C_i \left| i \right\rangle
 \qquad (C_i \; \textrm{real}) \ ,
\end{eqnarray}
where $| i \rangle$ is a CMPC (an eigenstate of $H_0$).

The angular momentum alignment for the state $| \Psi \rangle$ is
\begin{eqnarray}
\langle \Psi | J_x | \Psi \rangle = \sum_i C_i^2 \langle i | J_x | i
\rangle + 2\sum_{i<j}C_i C_j \langle i | J_x | j \rangle \ ,
\end{eqnarray}
and the kinematic MOI of state $| \psi \rangle$ is
\begin{eqnarray}
J^{(1)}=\frac{1}{\omega} \langle\Psi | J_x | \Psi \rangle \ .
\end{eqnarray}
Because $J_x$ is a one-body operator, the matrix element $\langle i | J_x | j \rangle$
($i\neq j$) may not vanish only when
$|i\rangle$ and $|j\rangle$ differ by
one particle occupation~\cite{Zeng1994_PRC50-1388}.
After a certain permutation of creation operators,
$|i\rangle$ and $|j\rangle$ can be recast into
\begin{eqnarray}
 |i\rangle=(-1)^{M_{i\mu}}|\mu\cdots \rangle \ , \qquad
|j\rangle=(-1)^{M_{j\nu}}|\nu\cdots \rangle \ ,
\end{eqnarray}
where $\mu$ and $\nu$ denote two different single-particle states,
and $(-1)^{M_{i\mu}}=\pm1$, $(-1)^{M_{j\nu}}=\pm1$ according to
whether the permutation is even or odd.
Therefore, the angular momentum alignment of
$|\Psi\rangle$ can be written as
\begin{eqnarray}
 \langle \Psi | J_x | \Psi \rangle = \sum_{\mu} j_x(\mu) + \sum_{\mu<\nu} j_x(\mu\nu)
 \ .
 \label{eq:jx}
\end{eqnarray}
where the diagonal contribution $j_x(\mu)$ and the
off-diagonal (interference) contribution $j_x(\mu\nu)$ can be written as
\begin{eqnarray}
j_x(\mu)&=&\langle\mu|j_{x}|\mu\rangle n_{\mu} \ ,
\\
j_x(\mu\nu)&=&2\langle\mu|j_{x}|\nu\rangle\sum_{i<j}(-1)^{M_{i\mu}+M_{j\nu}}C_{i}C_{j}
  \quad  (\mu\neq\nu) \ ,
\end{eqnarray}
and
\begin{eqnarray}
n_{\mu}=\sum_{i}|C_{i}|^{2}P_{i\mu} \ ,
\end{eqnarray}
is the occupation probability of the cranked orbital $|\mu\rangle$,
$P_{i\mu}=1$ if $|\mu\rangle$ is occupied in $|i\rangle$, and
$P_{i\mu}=0$ otherwise.

\section{Results and discussion}{\label{sec:resu}}

In this work, the deformation parameters $\varepsilon_2= 0.217$ and $\varepsilon_4=0$
are taken from Ref.~\cite{Bengtsson1986_ADNDT35-15},
which are chosen as an average of the neighboring even-even Hf and W isotopes.
The Nilsson parameters ($\kappa$ and $\mu$) are taken from Ref.~\cite{Bengtsson1985_NPA436-14},
and a slight change of $\kappa_6$ (modified from 0.062 to 0.068)
and $\mu_6$ (modified from 0.34 to 0.32) for neutron $N=6$ major shell is made.
The neutron Nilsson parameters are changed to account for the observed ground state of $^{167}$Hf.
In addition, the proton orbital $\pi 1/2^{-}[541]$ is slightly shifted upward
by about $0.2\hbar\omega_0$, which is adopted to avoid the defect caused by
the velocity-dependent $l^2$ term in the Nilsson potential for the MOIs
and alignments at the high-spin region~\cite{Andersson1976_NPA268-205}.
The effective pairing strengths can be determined by
the odd-even differences in nuclear binding energies,
and are connected with the dimension of the truncated CMPC space.
In this work, the CMPC space is constructed in the proton $N=4, 5$ major shell
and the neutron $N=5, 6$ major shell
with the truncation energies about 0.7$\hbar\omega_0$ both for protons and neutrons.
For $^{168}$Ta, $\hbar\omega_{\rm 0p}=7.106$~MeV
for protons and $\hbar\omega_{\rm 0n}=7.755$~MeV
for neutrons, respectively~\cite{Nilsson1969_NPA131-1}.
The dimensions of the CMPC space are
1000 for both protons and neutrons in the present calculation.
The corresponding effective pairing strengths are
$G_{\rm p}=0.34$~MeV for protons and $G_{\rm n}=0.46$~MeV for neutrons, respectively,
which are the same as those adopted in $^{166}$Ta~\cite{Zhang2016_SciChinaPMA59-672012}.

\begin{figure}[h]
\centering
\includegraphics[width=0.7\columnwidth]{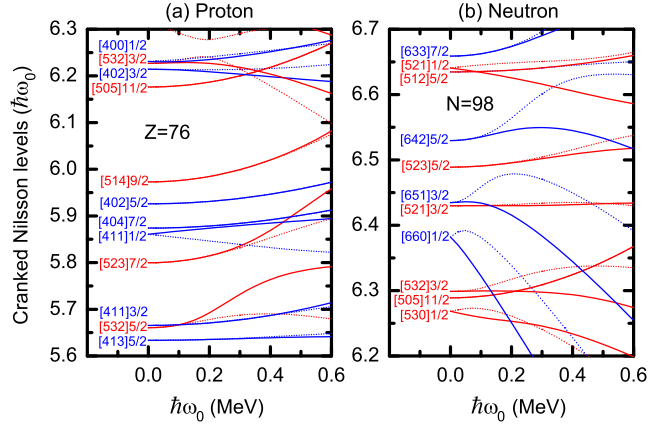}
\caption{\label{fig1} (Color online)
The cranked Nilsson levels near the Fermi surface of $^{168}$Ta
(a) for protons and (b) for neutrons.
The positive (negative) parity levels are denoted by blue (red) lines.
The signature $\alpha=+1/2$ ($\alpha=-1/2$) levels are denoted by solid (dotted) lines.
The deformation parameters $\varepsilon_2= 0.217$ and $\varepsilon_4=0$
are taken from Ref.~\cite{Bengtsson1986_ADNDT35-15},
which are taken as an average of the neighboring even-even Hf and W isotopes.
The Nilsson parameters ($\kappa$ and $\mu$) are taken from Ref.~\cite{Bengtsson1985_NPA436-14},
and a slight change of $\kappa_6$ (modified from 0.62 to 0.68)
and $\mu_6$ (modified from 0.34 to 0.32) for neutron $N=6$ major shell is made.
In addition, the proton orbital $\pi 1/2^{-}[541]$ is shifted upward by $0.2\hbar\omega_0$.
}
\end{figure}

Figure~\ref{fig1} shows the cranked Nilsson levels near the Fermi surface of $^{168}$Ta
(a) for protons and (b) for neutrons.
The positive (negative) parity levels are denoted by blue (red) lines.
The signature $\alpha=+1/2$ ($\alpha=-1/2$) levels are denoted by solid (dotted) lines.
Note that there are several high-$K$ orbitals close to the proton and neutron Fermi surfaces,
{\it e.g.}, $\pi 5/2^+[402]$, $\pi 7/2^+[404]$, $\pi 9/2^-[514]$,
$\nu 5/2^-[523]$ and $\nu 5/2^+[642]$.
Therefore, this may lead to the formation of various high-$K$ 2-quasiparticle isomers.
It also can be seen that near the Fermi surface, there exists a proton sub-shell at $Z=76$
and a neutron sub-shell at $N=98$.

\begin{figure}[!]
\centering
\includegraphics[width=0.6\columnwidth]{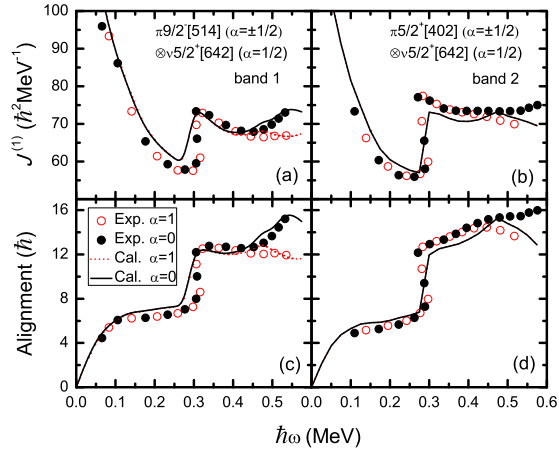}
\caption{\label{fig2} (Color online)
The experimental~\cite{Wang2010_PRC82-034315} and calculated
kinematic MOIs $J^{(1)}$ (upper panel) and alignments $i$ (lower panel)
of two high-spin rotational bands in $^{168}$Ta.
The alignment is defined as $i= \langle
J_x \rangle -\omega J_0 -\omega ^ 3 J_1$,
and the Harris parameters $J_0 = 28\ \hbar^2$MeV$^{-1}$ and $J_1 = 58\ \hbar^4$MeV$^{-3}$
are taken from Ref.~\protect\cite{Wang2010_PRC82-034315}.
The experimental MOIs and alignments
are denoted by black solid circles (signature $\alpha=0)$
and red open circles (signature $\alpha=1)$, respectively.
The calculated MOIs and alignments are
denoted by black solid lines (signature $\alpha=0$)
and red dotted lines (signature $\alpha=1$), respectively.
}
\end{figure}

Figure~\ref{fig2} shows the experimental~\cite{Wang2010_PRC82-034315} and calculated
kinematic MOIs $J^{(1)}$ (upper panel) and alignments $i$ (lower panel)
of two high-spin rotational bands in $^{168}$Ta.
The alignment is defined as $i= \langle
J_x \rangle -\omega J_0 -\omega^ 3 J_1$, and the Harris parameters
$J_0 = 28\ \hbar^2$MeV$^{-1}$ and $J_1 = 58\ \hbar^4$MeV$^{-3}$
are taken from Ref.~\protect\cite{Wang2010_PRC82-034315}.
The experimental MOIs and alignments
are denoted by black solid circles (signature $\alpha=0)$
and red open circles (signature $\alpha=1)$, respectively.
The calculated MOIs and alignments are
denoted by black solid lines (signature $\alpha=0$)
and red dotted lines (signature $\alpha=1$), respectively.
In Ref.~\cite{Theine1992_NPA536-418}, these two bands were firstly observed
and based on the relative intensity,
their configurations were tentatively assigned as
$\pi 9/2^-[514] \otimes \nu 5/2^+[642]$ ($\pi h_{11/2} \otimes \nu i_{13/2}$) for band 1 and
$\pi 5/2^+[402] \otimes \nu 5/2^+[642]$ ($\pi d_{5/2}  \otimes \nu i_{13/2}$) for band 2,
respectively.
Later on, these two bands were extended up to spin $\sim 40\hbar$ in Ref.~\cite{Wang2010_PRC82-034315},
in which the second upbending at $\hbar\omega\sim 0.5$~MeV
in ${}^{168}$Ta has been observed for the first time.
Then by analyzing the energy staggerings, electromagnetic transition probabilities
and rotational alignments of these two bands, the configurations were confirmed.
It can be seen in Fig.~\ref{fig2} that the signature splittings
at low rotational frequency region in these two bands are quite small.
Due to the large signature splitting of the $\nu i_{13/2}$ orbital,
these two bands should be constructed by the favored $\alpha = +1/2$ sequence of $\nu i_{13/2}$
coupled to both signature partners of the single proton.
Using these configuration assignments, the MOIs and alignments for these two bands are
reproduced quite well by the PNC-CSM calculations, which in turn supports the
configuration assignments~\cite{Wang2010_PRC82-034315, Theine1992_NPA536-418}.
Especially, the second upbending around $\hbar\omega\sim 0.5$~MeV
in the $\alpha=0$ sequence of band 1 is reproduced quite well and the
corresponding signature splitting is also reproduced.
Note that in Ref.~\cite{Wang2010_PRC82-034315}, the CSM can not reproduce this upbending,
which was tentatively interpreted as the crossing with proton
$\pi^2 9/2^-[514](\alpha=1/2) \otimes 1/2^-[541] (\alpha=-1/2)$ configuration.
However, the gradual upbending in the $\alpha=0$ sequence of band 2 at $\hbar\omega\sim 0.5$~MeV
is not reproduced by the PNC-CSM, which may need further investigation.
In the following, the level crossings in the backbending and upbending will be discussed in detail.

\begin{figure}[!]
\centering
\includegraphics[width=0.45\columnwidth]{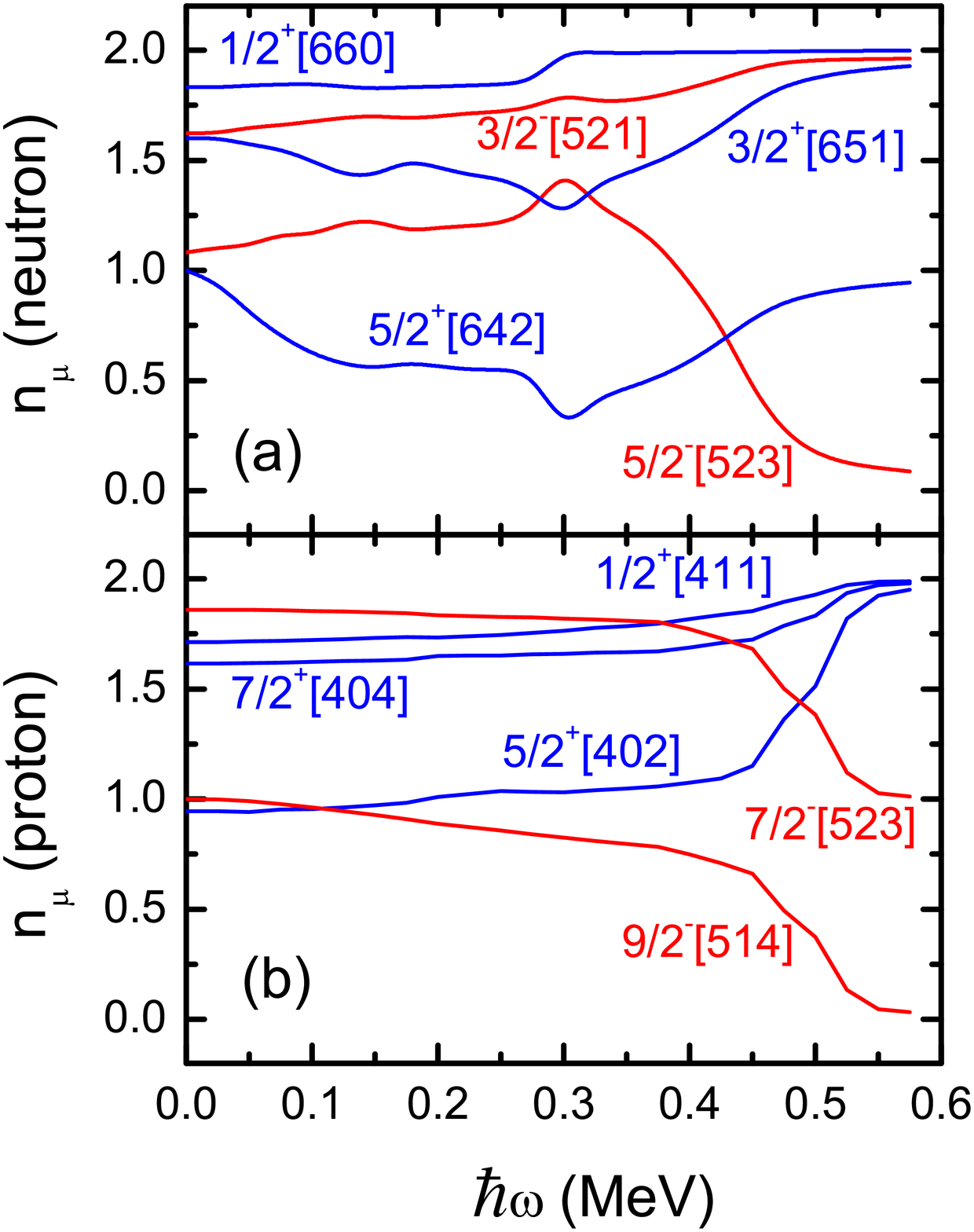}
\caption{\label{fig3} (Color online)
Occupation probability $n_\mu$ of each orbital
$\mu$ (including both $\alpha=\pm1/2$) near the Fermi surface
of band 1 in $^{168}$Ta.
The top and bottom rows are for neutrons and protons, respectively.
The positive (negative) parity levels are denoted by blue (red) lines.
The Nilsson levels far above ($n_{\mu}\sim0$) and far below ($n_{\mu}\sim2$)
the Fermi surface are not shown.
}
\end{figure}

\begin{figure}[!]
\centering
\includegraphics[width=0.45\columnwidth]{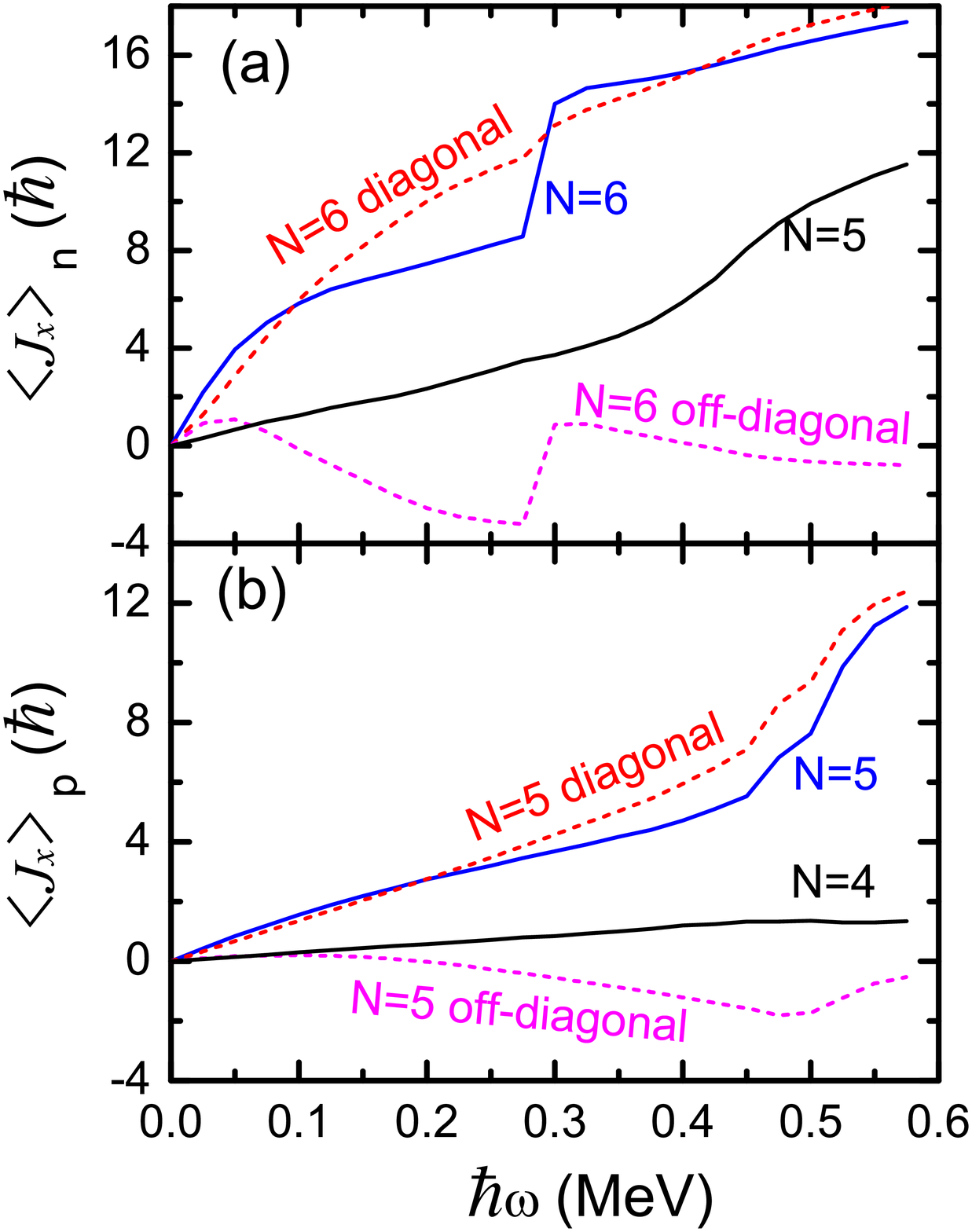}
\caption{\label{fig4} (Color online)
Contributions of (a) neutron $N=5, 6$ major shell and (b) proton $N=4, 5$ major shell
to the angular momentum alignment $\langle J_x\rangle$ for band 1 in $^{168}$Ta.
The contributions of diagonal $\sum_{\mu} j_x(\mu)$ and off-diagonal part
$\sum_{\mu<\nu} j_x(\mu\nu)$ in Eq.~(\protect\ref{eq:jx})
from the neutron $N=6$ and proton $N=5$ major shell are shown as dashed lines.
}
\end{figure}

It is well known that the backbending and upbending  phenomena in the rare-earth region
is caused by the alignment of the high-$j$ neutron $i_{13/2}$
and proton $h_{11/2}$ orbitals~\cite{Stephens1975_RMP47-43}.
As for band 1 of $^{168}$Ta, the proton $h_{11/2}$ orbital $\pi9/2^-[514]$
and the neutron $i_{13/2}$ orbital $\nu5/2^+[642]$ are all blocked.
Therefore, it is interesting to investigate the level crossings in this band.
To understand the first ($\hbar\omega \sim 0.3$~MeV) and
the second ($\hbar\omega \sim 0.5$~MeV) level crossings in band 1 of
$^{168}$Ta, the occupation probability $n_\mu$ of each orbital
$\mu$ (including both $\alpha=\pm1/2$) near the Fermi surface
of band 1 are shown in Fig.~\ref{fig3}.
The top and bottom rows are for neutrons and protons, respectively.
The positive (negative) parity levels are denoted by blue (red) lines.
The Nilsson levels far above ($n_{\mu}\sim0$) and far below ($n_{\mu}\sim2$)
the Fermi surface are not shown.
It can be seen in Fig.~\ref{fig3}(a) that for neutron,
the Coriolis mixing between the neutron $i_{13/2}$ orbitals
$\nu5/2^+[642]$ and $\nu3/2^+[651]$ is very strong since they are all
very close to the Fermi surface (see Fig.~\ref{fig1}).
Around the rotational frequency $\hbar\omega \sim 0.3$~MeV,
the occupation probabilities of $\nu5/2^+[642]$ and $\nu3/2^+[651]$
suddenly increase, while the occupation probability of $\nu5/2^-[523]$
suddenly decreases.
Therefore, the first backbending around $\hbar\omega \sim 0.3$~MeV in band 1
may come from the contribution of these two $i_{13/2}$ neutrons.
At $\hbar\omega > 0.3$~MeV, the occupation probabilities for all neutron orbitals
change gradually, so they may not contribute to the second upbending.
Fig.~\ref{fig3}(b) shows the proton occupation probabilities of band 1.
It can be seen that around $\hbar\omega \sim 0.5$~MeV,
the occupation probabilities of $\pi7/2^-[523]$ and $\pi9/2^-[514]$
decrease, while the occupation probability of $\pi5/2^+[402]$ increases.
Therefore, the second upbending around $\hbar\omega \sim 0.5$~MeV in band 1
may come from the contribution of these two $h_{11/2}$ protons.

The contributions of neutron $N=5, 6$ major shell (upper panel) and proton $N=4, 5$ major shell (lower panel)
to the angular momentum alignment $\langle J_x\rangle$ for band 1 in $^{168}$Ta are shown in Fig.~\ref{fig4}.
The contributions of diagonal $\sum_{\mu} j_x(\mu)$ and off-diagonal part
$\sum_{\mu<\nu} j_x(\mu\nu)$ in Eq.~(\protect\ref{eq:jx})
from the neutron $N=6$ and proton $N=5$ major shell are shown as dashed lines.
It should be noted that in this figure, the smoothly increasing part of the
angular momentum alignment represented by the Harris formula is not
subtracted ({\it cf.} the caption of Fig.~\ref{fig2}).
It can be seen in Fig.~\ref{fig4}(a) that, the first backbending at $\hbar\omega \sim 0.3$~MeV
mainly comes from the contribution of neutron $N=6$ major shell.
Furthermore, this backbending mainly comes from the off-diagnoal part of $N=6$ major shell.
Fig.~\ref{fig4}(b) shows that, the second upbending at $\hbar\omega \sim 0.5$~MeV
mainly comes from the contribution of proton $N=5$ major shell.
Furthermore, this upbending mainly comes from the diagnoal part of $N=5$ major shell.
Meanwhile, the off-diagnoal part also contribute a little.

\begin{figure}[h]
\centering
\includegraphics[width=0.45\columnwidth]{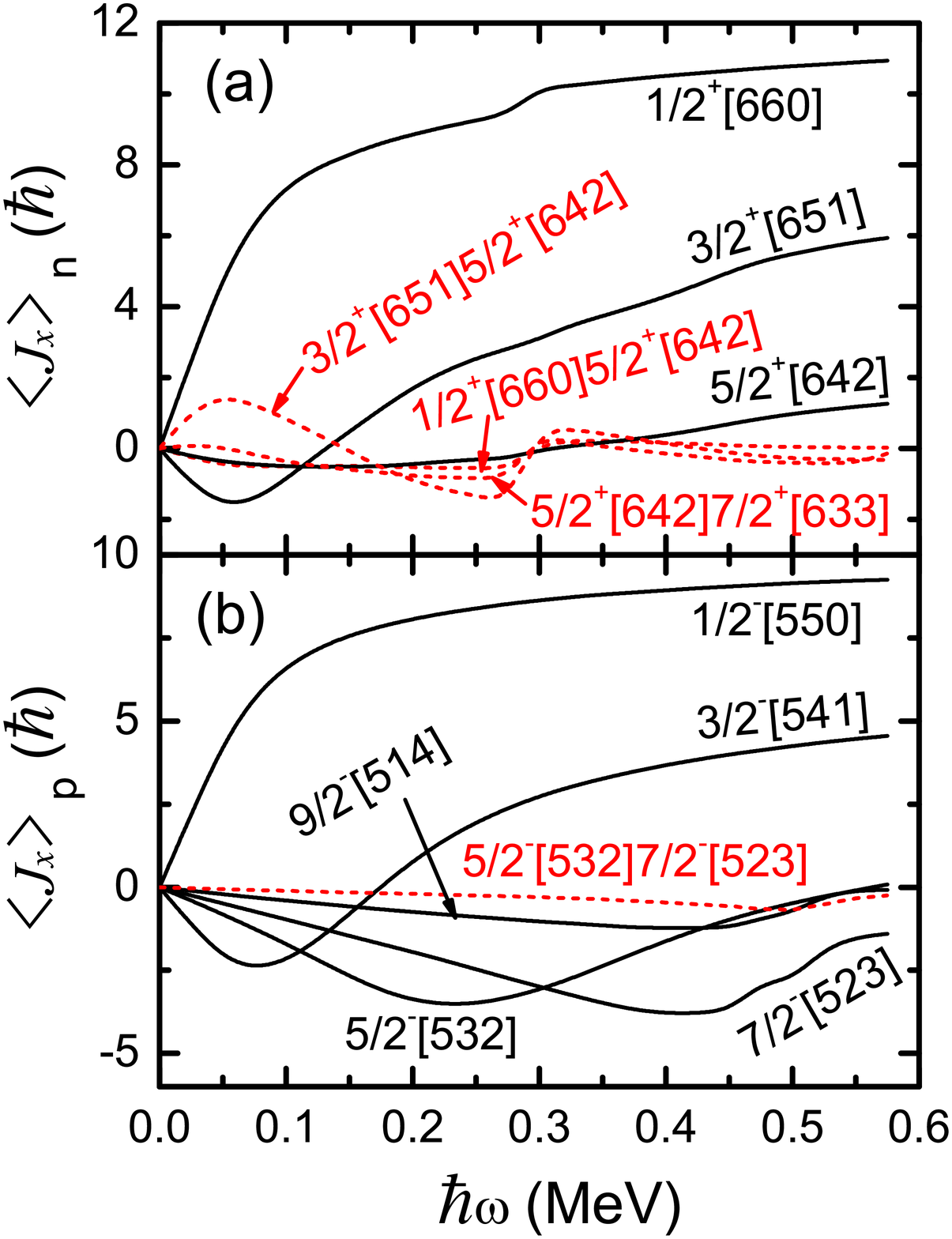}
\caption{\label{fig5} (Color online)
Contribution of (a) each neutron orbital in the  $N=6$
major shell and (b) each proton orbital in the $N=5$
major shell to the angular momentum alignments $\langle J_x\rangle$
of band 1 in $^{168}$Ta.
The diagonal (off-diagonal) part $j_x(\mu)$ [$j_x(\mu\nu)$] in
Eq.~(\protect\ref{eq:jx}) is denoted by black solid (red dotted) lines.
}
\end{figure}

In order to have a  more clear understanding of the level crossing mechanism,
the contribution of (a) each neutron orbital in the  $N=6$
major shell and (b) each proton orbital in the $N=5$
major shell to the angular momentum alignments $\langle J_x\rangle$
of band 1 in $^{168}$Ta are shown in Fig.\ref{fig5}(a).
The diagonal (off-diagonal) part $j_x(\mu)$ [$j_x(\mu\nu)$] in
Eq.~(\protect\ref{eq:jx}) is denoted by black solid (red dotted) lines.
It can be seen from Fig.~\ref{fig5} that for neutron, the off-diagonal part
$j_x\left(\nu 3/2^+[651] \nu 5/2^+[642]\right)$ changes a lot at $\hbar\omega \sim 0.3$~MeV.
The alignment gain after the backbending mainly comes from this interference term.
In addition, the off-diagonal parts
$j_x\left(\nu 1/2^+[660] \nu 3/2^+[651]\right)$ and
$j_x\left(\nu 5/2^+[642] \nu 7/2^+[633]\right)$ also have obvious contributions.
This tells us that the first backbending at $\hbar\omega \sim 0.3$~MeV is mainly caused
by the neutron $i_{13/2}$ orbitals.
From Fig.~\ref{fig5}(b) one can see that for proton,
the diagonal part $j_x\left(\pi 7/2^-[523]\right)$ contributes a lot to the upbending.
The diagonal part $j_x\left(\pi 9/2^-[514]\right)$ and off-diagonal part
$j_x\left(\pi 5/2^-[532] \pi 7/2^-[523]\right)$ also contribute a little.
Therefore, we can get that the second upbending at $\hbar\omega \sim 0.5$~MeV is
mainly caused by the proton $h_{11/2}$ orbitals.

\section{Summary}{\label{sec:summ}} \vspace*{-1mm}

The structures of two observed 2-quasiparticle high-spin rotational bands in the
doubly-odd nucleus ${}^{168}$Ta are investigated using the cranked shell model
with pairing correlations treated by a particle-number conserving method,
in which the blocking effects are taken into account exactly.
The experimental moments of inertia and alignments
are reproduced very well by the particle-number conserving calculations,
which confirms the configuration assignments for these two bands in previous works.
The backbending and upbending mechanisms in these two bands are analyzed by calculating the
occupation probabilities of each orbital close to the Fermi surface
and the contributions of each orbital to the total angular momentum alignments.
It was found that for the first backbending at $\hbar\omega \sim 0.3$~MeV,
the interference terms between the neutron $\nu i_{13/2}$ orbitals contribute a lot.
For the second upbending at $\hbar\omega \sim 0.5$~MeV,
the diagonal part of proton $h_{11/2}$ orbitals contributes a lot,
and the off-diagonal part of proton $h_{11/2}$ orbitals also contributes a little.
Note that the second gradual upbending in band 2 with signature $\alpha=0$
is not reproduced by the present calculation.
This may be caused by the deformation changing with increasing rotational frequency,
especially the triaxial deformation.
However, in the present PNC-CSM, the deformation is fixed.
Therefore, more sophisticated theory is needed to perform further investigations for this band.
The recently developed shell-model-like approach, originally referred to as PNC method, based on the
cranking covariant density functional theory can treat
the deformation self-consistently with increasing rotational
frequency and may provide more detailed information for this upbending.

\section{Acknowledgement}
\label{acknowledgement}
This work was partly supported by the National Natural Science
Foundation of China (Grants No. 11505058, 11775112, 11775026),
and the Fundamental Research Funds for the Central Universities (2018MS058).


\end{document}